\documentclass[3p,times]{elsarticle}

\usepackage{ecrc}

\usepackage{epstopdf}

\volume{00}

\firstpage{1}

\journalname{Nuclear Physics A}

\runauth{L. Yi}

\jid{nupha}

\jnltitlelogo{Nuclear Physics A}

\usepackage{graphicx}
\usepackage{amsmath,amssymb}
\usepackage{lineno}

\begin{document}

\begin{frontmatter}



\title{Search for ``Ridge" in d+Au Collisions at RHIC by STAR}

%% Single author (and collaboration) - please insert
\author{Li Yi (for the STAR\fnref{col1} Collaboration)}
\fntext[col1] {A list of members of the STAR Collaboration and acknowledgements can be found at the end of this issue.}
\address{Department of Physics and Astronomy, Purdue University, West Lafayette, IN, 47907, USA}

%% For multiple authors, replace the above by:

%\author[label1]{Author1}
%\author[label2]{Author2}

%\address[label1]{Address 1}
%\address[label2]{Address 2}

\begin{abstract}
Dihadron correlations are measured in d+Au collisions at 200 GeV by the STAR detector. The correlated yields with uniform background subtraction are studied in high- and low-multiplicity collisions. The effects of multiplicity selection bias on jet-like correlations are discussed. Finite correlated yields are observed on the near-side at large pseudo-rapidity separation in high-multiplicity collisions.
%A template for preparing contributions to the proceedings of Quark Matter 2014. The file should be compiled
%with {\em pdflatex}. Figures can be pdf or eps, as illustrated in Fig.~\ref{fig:generic}. If the conversion eps $\to$ pdf
%does not work automatically on your system, you can convert eps files to pdf using a tool like {\em epstopdf}. For special options see\\ 
%\verb!http://www.elsevier.com/author-schemas/preparing-crc-journal-articles-with-latex!.
\end{abstract}


\end{frontmatter}

%%\linenumbers


\section{Introduction}
\label{sec:intro}
Long-range pseudo-rapidity separation ($\Delta\eta$) correlations at small azimuthal difference ($\Delta\phi$), called the ridge, have been observed in high-multiplicity p+p and p+Pb collisions at the LHC \cite{CMS1,CMS2,ALICE,ATLAS}. A subtraction of the dihadron correlation in low-multiplicity p+Pb collisions from high-multiplicity ones reveals a back-to-back double-ridge structure (at $\Delta\phi \approx 0$ and $\pi$) \cite{ALICE,ATLAS}. A similar double-ridge is also observed by PHENIX in d+Au collisions using the same subtraction method \cite{PHENIX1}. Differences between multiplicity-selected d+Au collisions (and p+p collisions) have been observed before by STAR \cite{STARdAu}. Recent ALICE results show a mass ordering of the proton and pion elliptic anisotropy parameters, $v_{2}$, characterizing the double-ridge correlations in p+Pb collisions \cite{ALICEmassordering}. A similar mass ordering is observed by PHENIX in d+Au collisions \cite{PHENIXmassordering}. The hydrodynamic models, where collective flow develops in p/d+A collisions \cite{Bozek, Bzdak, Qin}, and the Color Glass Condensate model, where two-gluon production is enhanced, provide two possible explanations to the ridge in small systems \cite{Dusling}. Measurements from the large acceptance STAR detector should shed light on the long-range correlation in d+Au collisions at RHIC.

%Mauris lacinia lorem sit amet ipsum. Nunc quis urna dictum turpis accumsan semper~\cite{ref1}
%\begin{equation}
%\bar{x}=\frac{1}{n}\sum_{i=1}^{n}x_{i}
%\end{equation}
%\begin{equation}
%\int_{0}^{\infty}e^{-\alpha x^{2}}=\frac12\sqrt{\frac{\pi}{\alpha}}
%\end{equation}

%\newpage
%%{\bf  A generic figure, in pdf and eps, is shown in Fig.~\ref{fig:generic}.}
%\vspace*{1cm}

%\begin{figure}
%\begin{center}
%\includegraphics*[width=9.cm]{bessel1}\\
%\includegraphics*[width=9.cm]{bessel2}
%\caption{
%Generic figures, top is pdf, bottom is eps}
%\label{fig:generic}
%\end{center}
%\end{figure}

\section{Event and track selections}
\label{sec:dataset}
The data used in this analysis are composed of 4 million d+Au events at nucleon-nucleon center-of-mass energy of $\sqrt{s_{\rm{NN}}} = 200$ GeV, collected by the STAR detector in 2003. Three main detectors used in this analysis are the Time Projection Chamber (TPC), the Forward Time Projection Chamber (FTPC), and the Zero Degree Calorimeter (ZDC). The minimum bias events were triggered by the ZDC in the Au beam direction (ZDC-Au). The events are required to have the reconstructed
primary vertex within 50 $cm$ of the TPC center along the beam direction. The charged tracks reconstructed in the TPC and FTPC are required to satisfy the following conditions: the distance of the closest approach to the primary vertex less than 3 $cm$ to remove secondary tracks from particle decays; the number of fit points greater than 25 (5) in the TPC (FTPC) for good track reconstruction, and larger than 0.51 times the maximum number of possible fit points to avoid split
tracks. The track pseudo-rapidity cuts are $|\eta|<1$ ($2.8<|\eta|<3.8$) in the TPC (FTPC) \cite{STAR:dAu4suppression}.

\section{Data analysis}
\label{sec:ana}
Two sets of dihadron correlations are analyzed in this study: TPC-TPC correlations with both the trigger and associated particles from the TPC ($-2<\Delta\eta<2$), and TPC-FTPC correlations with the trigger particle from the TPC and the associated particle from the FTPC ($2<|\Delta\eta|<4.5$). Both trigger and associated particles have transverse momentum $1<p_{T}<3$ GeV/$c$. The $\Delta\eta$-$\Delta\phi$ dihadron correlation ($\Delta\eta=\eta_{\text{assoc}}-\eta_{\text{trig}}$ and $\Delta\phi=\phi_{\text{assoc}}-\phi_{\text{trig}}$) is given by
\begin{equation}
\frac{1}{N_{trig}} \frac{d^{2} N}{ d \Delta\eta d \Delta\phi} = \frac{1}{N_{trig}} \frac{S(\Delta\eta,\Delta\phi)/\epsilon_{\text{assoc}}}{B(\Delta\eta,\Delta\phi)/\langle B(\Delta\eta|_{\text{100\%}},\Delta\phi)\rangle}.
\label{eqn:dihadron}
\end{equation}
Here $S=\frac{1}{N_{\text{trig}}}\frac{d^{2} N^{\text{same}}}{d\Delta\eta d\Delta\phi}$ is the raw dihadron correlation for pairs in the same event; and $B=\frac{1}{N_{\text{trig}}}\frac{d^{2} N^{\text{mix}}}{d\Delta\eta d\Delta\phi}$ is for trigger and associated particles from different (mixed) events. The mixed event background serves as the correction for the detector two-particle acceptance; $\langle B \rangle$ is the average $B$ at fixed $\Delta\eta|_{\text{100\%}}$, where the two-particle acceptance is 100\%; $|\Delta\eta|_{\text{100\%}}=0$ ($3.3$) for TPC-TPC (TPC-FTPC) correlations. The mixed events are required to have primary vertices within 1 $cm$ of each other in the beam direction to resemble similar acceptance, and have similar event characteristics. 

The yields are corrected for the associated particle tracking efficiencies, $\epsilon_{\text{assoc}}=$ 85\% for TPC tracks and 70\% for FTPC tracks. The underlying event background is further subtracted by $\Delta\eta$-dependent Zero-Yield-At-Minimum (ZYAM) method \cite{ZYAM}. 

The systematic uncertainties are estimated by varying the width of the ZYAM normalization $\Delta\eta$ range from 0.4 to 0.2 and 0.6 radian. An additional 5\% tracking systematic uncertainty is applied.

%\section{Correlation yield result and discussion}
%\label{sec:result}
%The following discussion will be presented as the dihadron correlation projections on $\Delta\eta$ and $\Delta\phi$.
%
\section{Multiplicity selection bias}
\label{sec:bias}

High multiplicity is required for event selection in order to observed the ridge. In this analysis, we use the raw charged particle multiplicity in $-3.8<\eta<-2.8$ measured by the Au direction FTPC (FTPC-Au) for event selection, similar to the $\eta$ range used by PHENIX. We also use the neutral particle energy deposited in ZDC-Au. The correlation between the FTPC-Au multiplicity and the ZDC-Au neutral energy is positive but broad, thus these two measurements select significantly different event samples. In this contribution, the FTPC multiplicity selection is used for the TPC-TPC correlations, while ZDC energy selection for the TPC-FTPC correlations.

\begin{figure}[hbt]
\begin{center}
\includegraphics*[width=4.8cm]{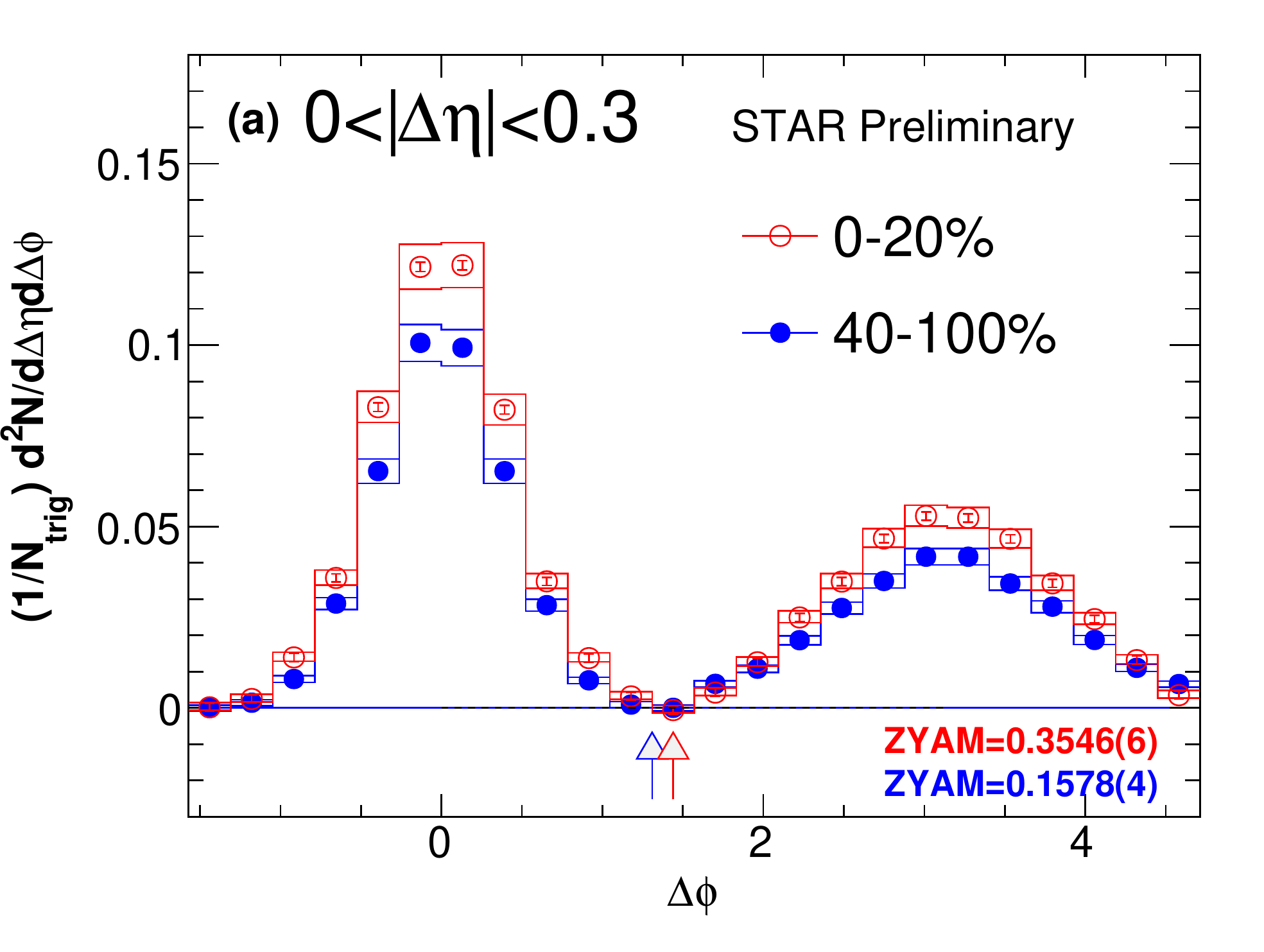}
\includegraphics*[width=4.8cm]{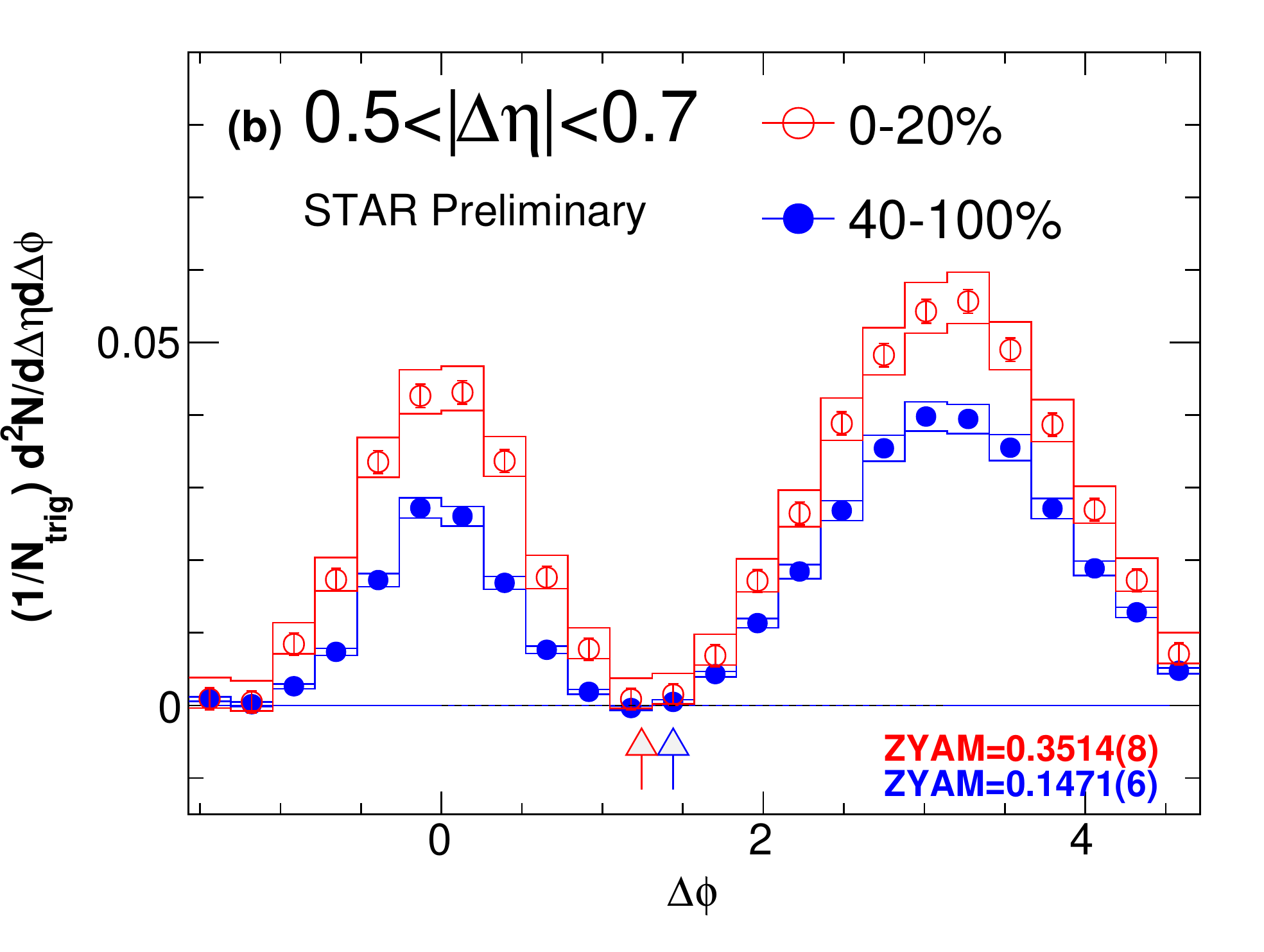}
\includegraphics*[width=4.8cm]{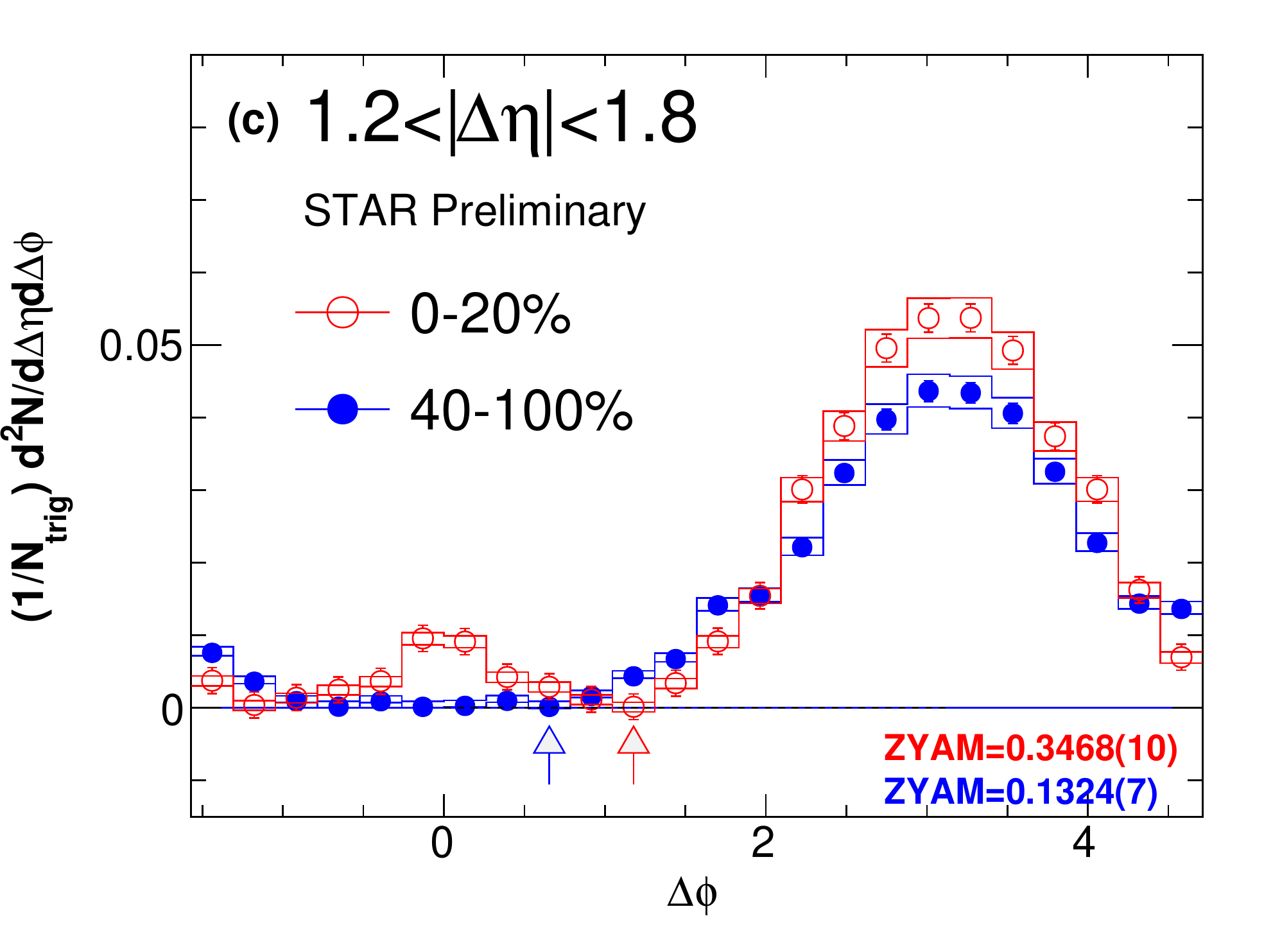}
\caption{Dihadron $\Delta\phi$ correlations for (a) $0<|\Delta\eta|<0.3$, (b) $0.5<|\Delta\eta|<0.7$ and (c) $1.2<|\Delta\eta|<1.8$ in d+Au collisions at $\sqrt{s_{\text{NN}}}=200$ GeV for charged particles of $1<p_{T}<3$ GeV/$c$. Both the trigger and associated particles are from the TPC. FTPC-Au multiplicity is used for event selection. The red open circles represent the high-multiplicity (0-20\%) collisions. The blue solid dots represent the low-multiplicity (40-100\%) collisions. The ZYAM backgrounds are listed on the plot.}
\label{fig:TPCDphi}
\end{center}
\end{figure}

Fig.~\ref{fig:TPCDphi} shows the TPC-TPC $\Delta\phi$ correlations at three $|\Delta\eta|$ ranges for high (0-20\%) and low (40-100\%) FTPC-Au multiplicity collisions. The near-side correlated yield in high-multiplicity collisions is larger than that in low-multiplicity ones. At large $|\Delta\eta|$, the high-multiplicity data have an excess correlated yield on the near-side ($|\Delta\phi|\approx0$) over low-multiplicity data, as shown in Fig.~\ref{fig:TPCDphi} (c). This will be discussed in the next section.

%\subsection{Away-side discussion: Jet-like correlation and multiplicity selection effect}
%\label{sec:awayside}

\begin{figure}[hbt]
\begin{center}
\includegraphics*[width=4.8cm]{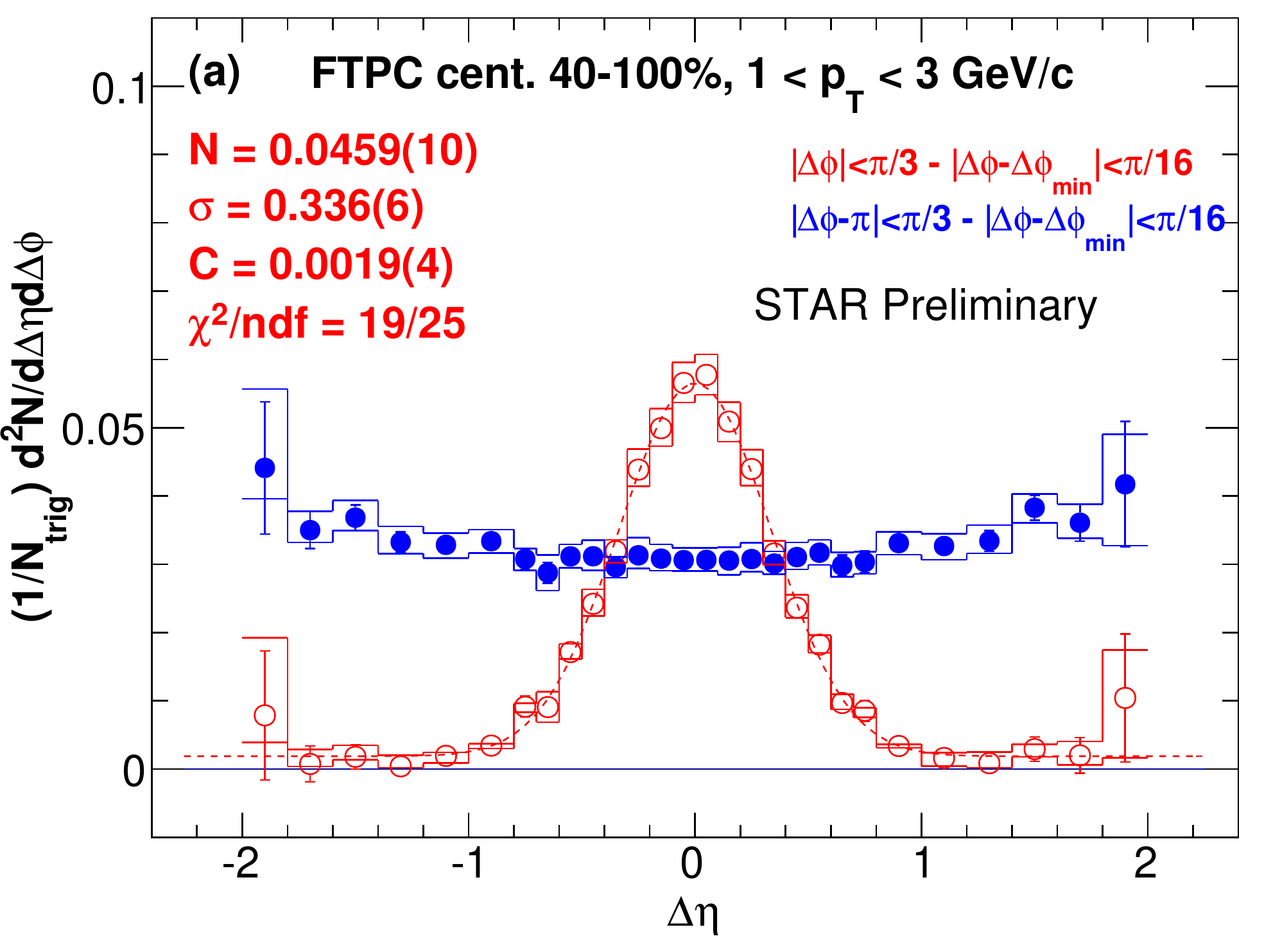}
\includegraphics*[width=4.8cm]{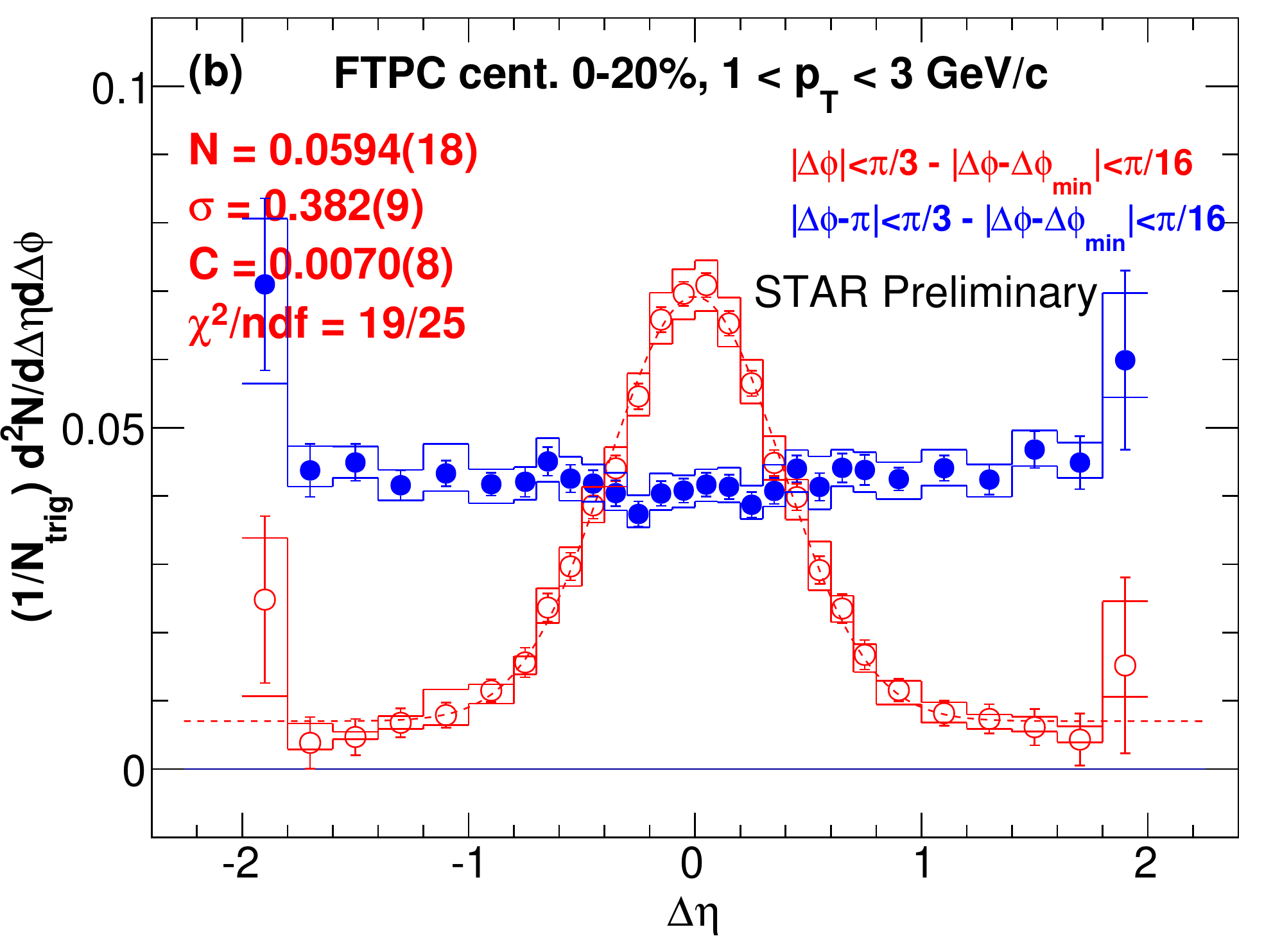}
\caption{Dihadron $\Delta\eta$ correlations in FTPC-Au multiplicity selected (a) 40-100\% and (b) 0-20\% d+Au collisions at $\sqrt{s_{\text{NN}}}=200$ GeV for charged particles of $1<p_{T}<3$ GeV/$c$. The red open points are near-side correlations, while the blue solid for away-side correlations. The Gausian$+$pedestal fit results are listed as N for the Guassian area, $\sigma$ for Gaussian width, $C$ for the pedestal constant.}
\label{fig:Deta}
\end{center}
\end{figure}

Fig.~\ref{fig:Deta} shows the TPC-TPC $\Delta\eta$ correlations in low- and high-multiplicity events. The correlations are larger in high- than low-multiplicity collisions. 

Dihadron correlations in d+Au collisions are dominated by jet-like correlations. To characterize jet-like correlations, the near-side correlations are fit with a Gaussian$+$constant pedestal. The fit results are shown in Fig.~\ref{fig:Deta}. The ratio of the high- and low-multiplicity Gaussian areas is found to be $\alpha=1.29\pm0.05$. This would represent the ratio of the jet-like correlated yields if the Gaussians represent jets (and the pedestals represent non-jet contributions). The non-unity $\alpha$ parameter suggests an event selection bias on the jet population. The high-multiplicity events appear to select jets with larger yield and wider $\Delta\eta$ distribution.

\begin{figure}[hbt]
\begin{center}
\includegraphics*[width=4.8cm]{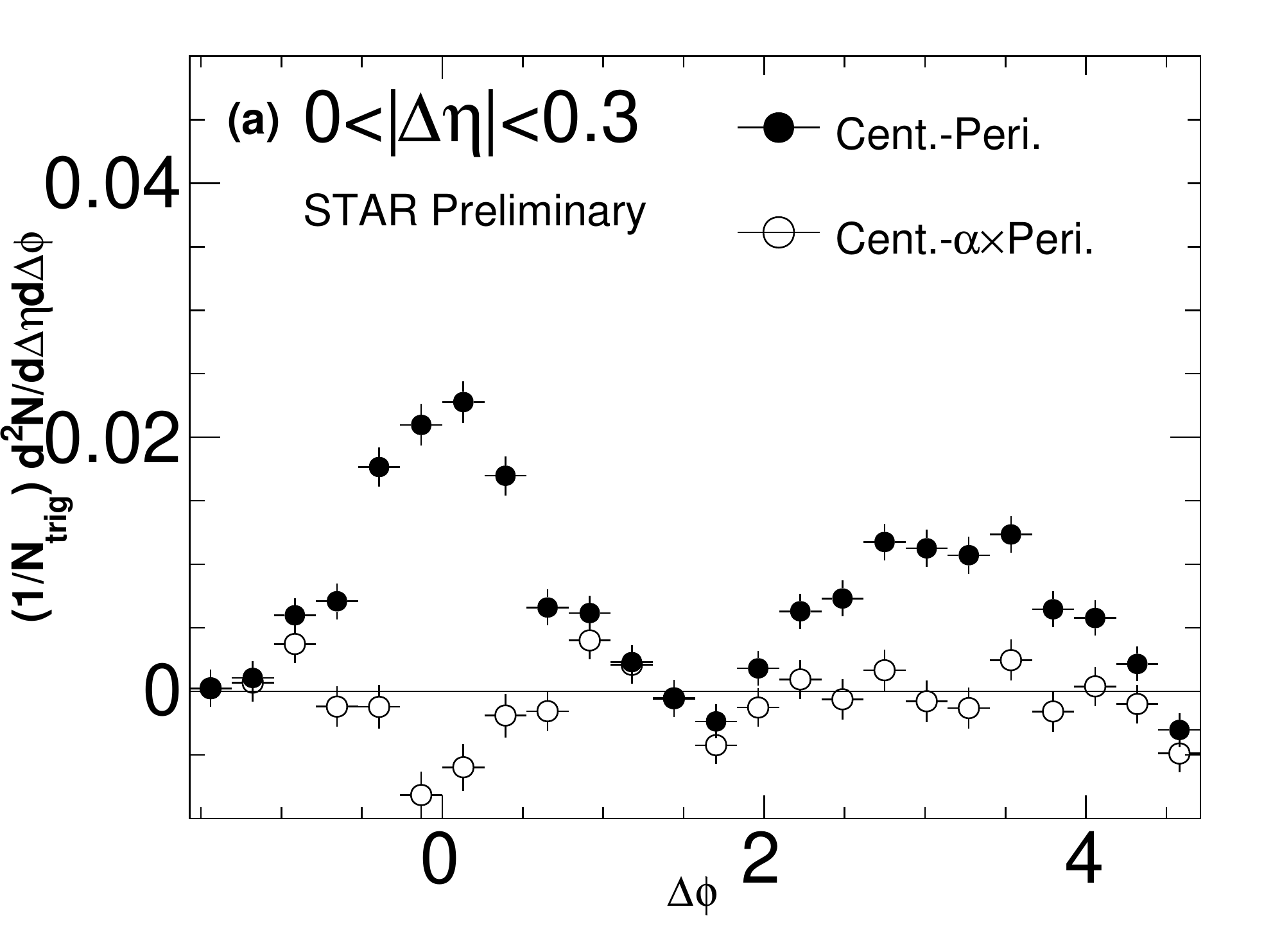}
\includegraphics*[width=4.8cm]{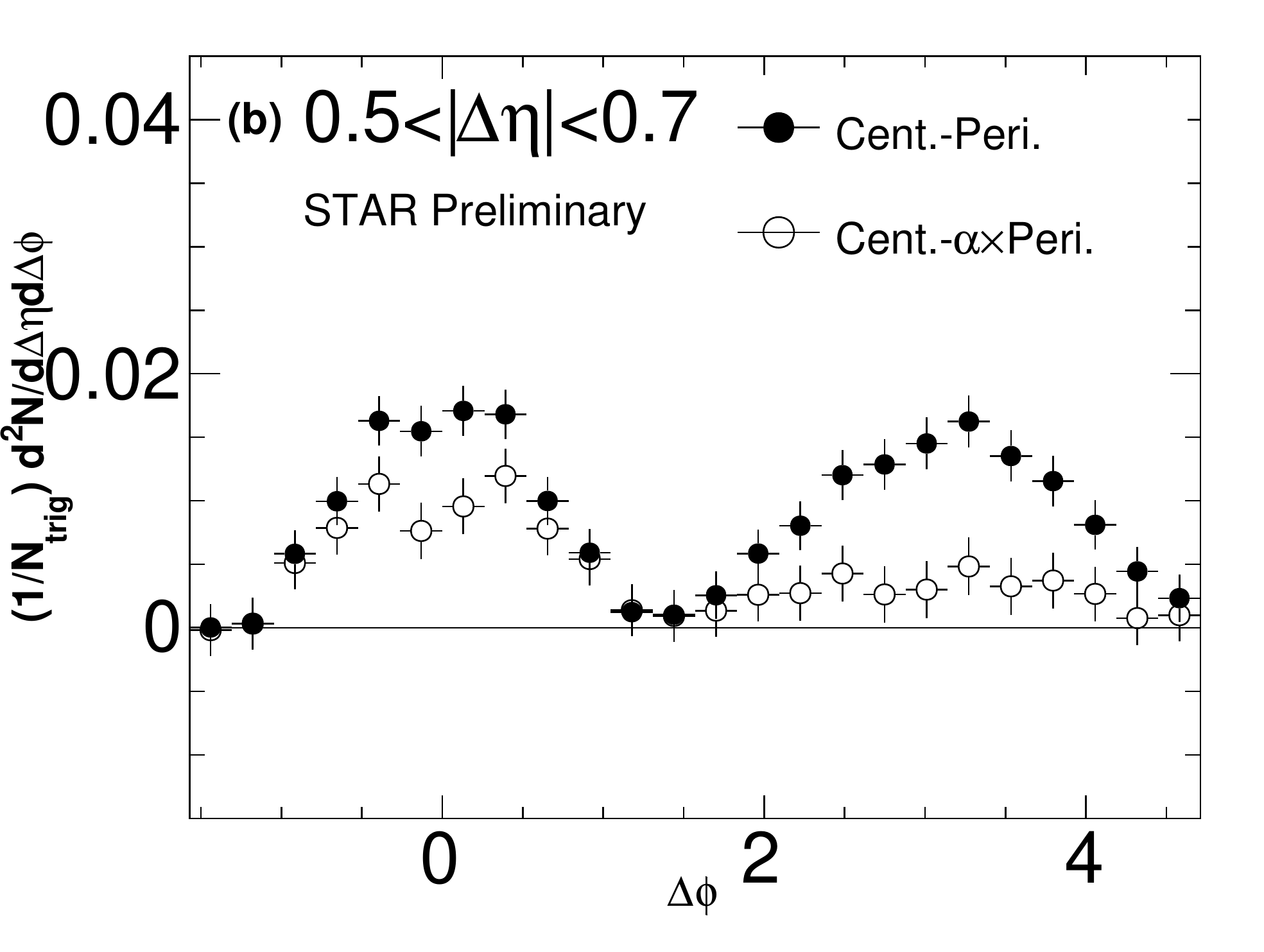}
\includegraphics*[width=4.8cm]{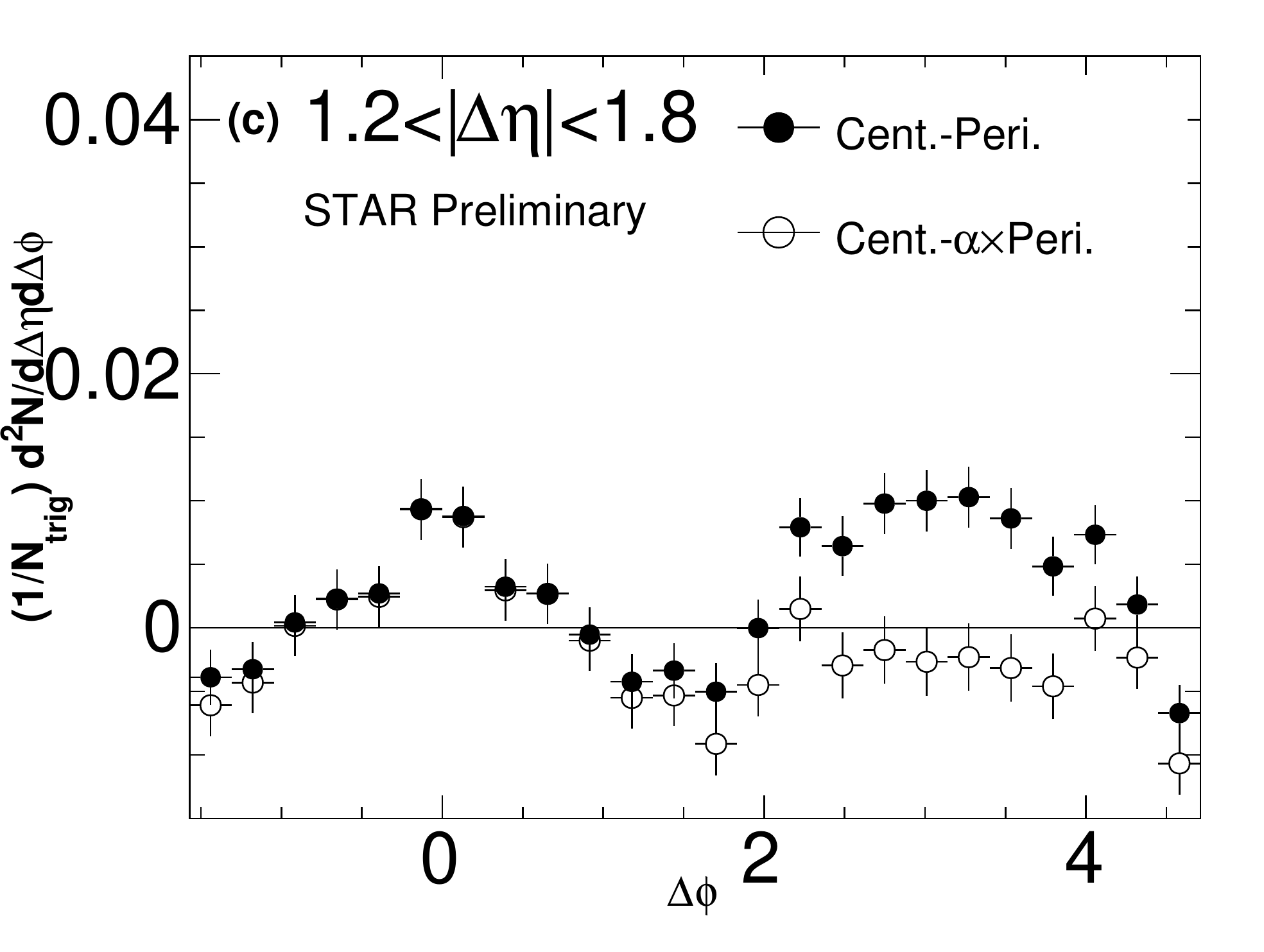}
\caption{Dihadron $\Delta\phi$ correlation difference between high- and low-multiplicity collisions in (a) $0<|\Delta\eta|<0.3$, (b) $0.5<|\Delta\eta|<0.7$ and (c) $1.2<|\Delta\eta|<1.8$ in d+Au collisions at $\sqrt{s_{\text{NN}}}=200$ GeV for charged particles of $1<p_{T}<3$ GeV/$c$. Both the trigger and associated particles are from the TPC. FTPC-Au multiplicity is used for event selection. The solid dots represent ``Cent.$-$Peri.". The open circles represent ``Cent.$-\alpha\times$Peri.", where $\alpha$ is near-side Gaussian area ratio in high- and low-multiplicity collisions. The error bars are statistical errors.}
\label{fig:cent_peri}
\end{center}
\end{figure}

Since the away-side jet spreads over a wide $\Delta\eta$, it cannot be isolated. Because of momentum conservation, the away-side correlated yield likely scales with the near-side yield. The open circles in Fig.~\ref{fig:cent_peri} represent the difference between high- and low-multiplicity events, with the latter first multiplied by the $\alpha$ parameter from the fit. This scaling would be a first order correction to the multiplicity selection bias on jet-like correlations, such that the away-side jet contributions would be subtracted. Indeed, the away-side yields are approximately zero for all $|\Delta\eta|$ ranges shown in Fig.~\ref{fig:cent_peri}. This suggests that the difference in the away-side long-range correlations between high- and low-multiplicity events is mostly from jet-like correlations.

The solid dots in Fig.~\ref{fig:cent_peri} show the simple difference between high- and low-multiplicity data, as done by PHENIX \cite{PHENIX1}. The peak magnitudes on the near-side and away-side turn out to be similar, resembling a double-ridge. As the large acceptance STAR data show, the resulting double-ridge structure may well be due to residual jet correlations which remain after the simple subtraction of the low- from the high-multiplicity correlation distributions. 
%As the STAR large acceptance data demonstrate, the double-ridge may be mostly a jet contribution.

%It is, however, possible that the near-side non-jet contribution is non-uniform in $\Delta\eta$, for example, peaking at $\Delta\eta=0$. Such a non-jet contribution could result in a relatively similar jet correlated yields. However, the resulant jet correlated $\Delta\eta$ distribution would be even wider in high-multiplicity compared with low-multiplicity collisions. Our conclusion is robust that the jet population is affected by the multiplicity selection. Using low-multiplicity data to subtract jet contributions in high-multiplicity collisions is highly questionable. 
From the different yields and widths shown in Fig.~\ref{fig:Deta} for the near-side $\Delta\eta$ dihadron correlations, we conclude that the jet population is affected by the multiplicity selection of the low- and high-multiplicity event classes. Using low-multiplicity data to subtract the jet contributions from the high multiplicity event dihadron correlations is thus only a rough approximation to the desired correction. The interpretation of the double-ridge structure resulting from such a subtraction in d+Au collisions at RHIC should be taken with caution.

\section{Near-side long range correlation}
\begin{figure}[hbt]
\begin{center}
\includegraphics*[width=4.8cm]{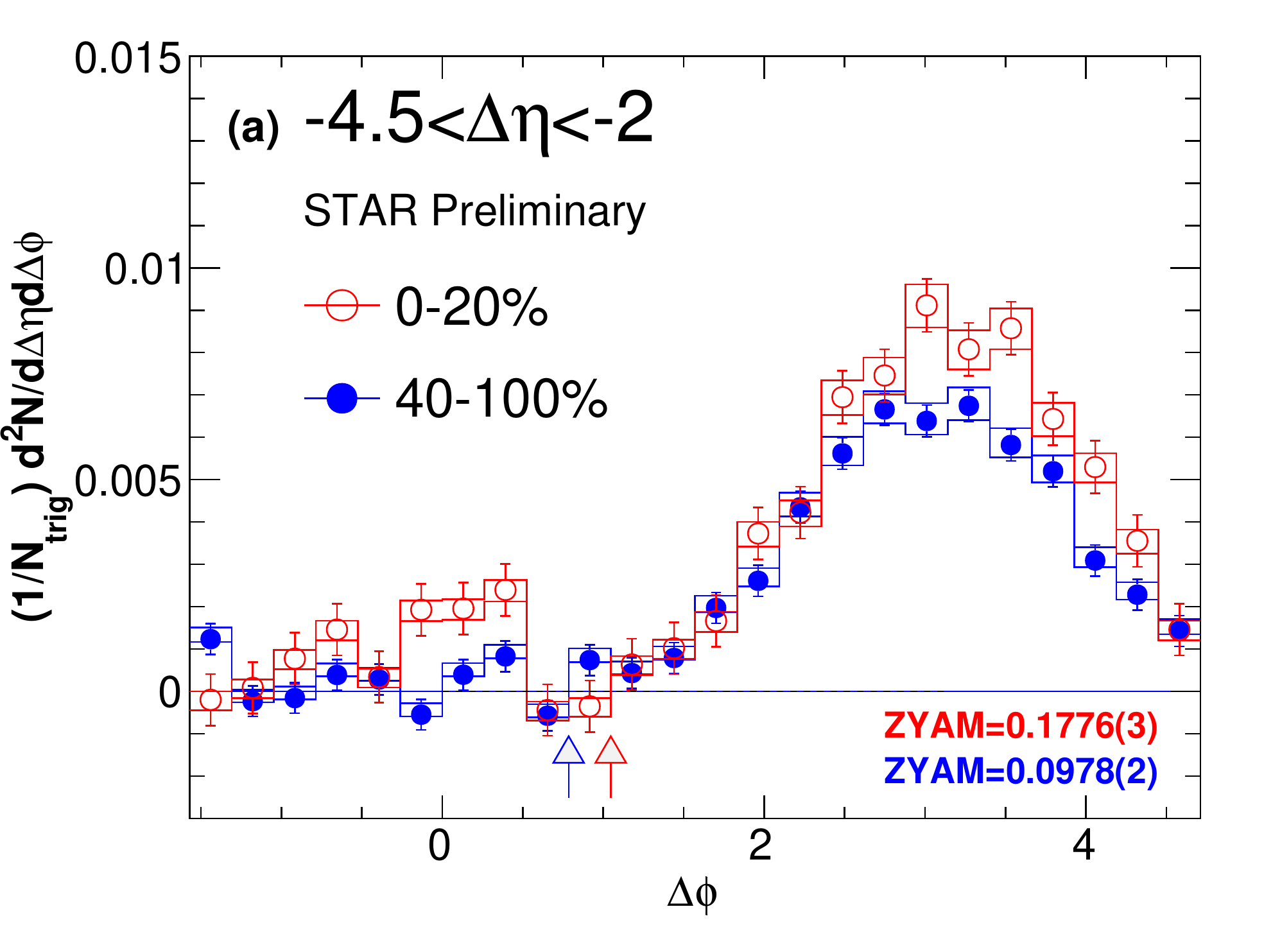}
\includegraphics*[width=4.8cm]{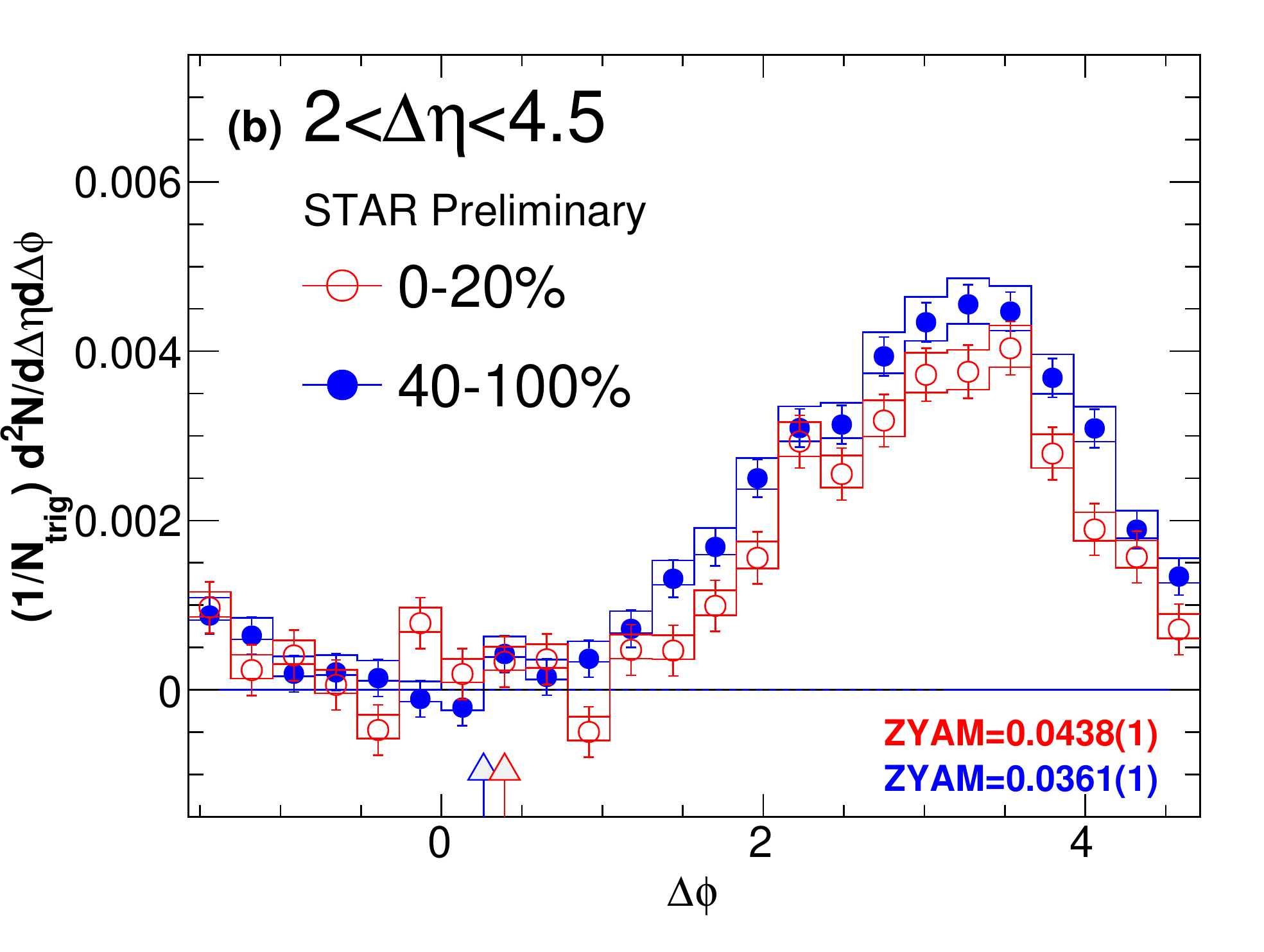}
\caption{Dihadron $\Delta\phi$ correlations in (a) $-4.5<\Delta\eta<-2.5$ and (b) $2.5<\Delta\eta<4.5$ in d+Au collisions at $\sqrt{s_{\text{NN}}}=200$ GeV for charged particles of $1<p_{T}<3$ GeV/$c$. The trigger particle is from TPC and the associated particle from FTPC. ZDC-Au neutral energy is used for event selection. The red open circles represent the high-multiplicity (0-20\%) collisions. The blue solid dots represent the low-mulitiplicity (40-100\%) collisions. The subtracted background values are listed on the plot.}
\label{fig:FTPCDphi}
\end{center}
\end{figure}

As seen in Fig.~\ref{fig:TPCDphi} (c), the TPC-TPC correlation shows a near-side peak at $|\Delta\eta|\approx1.5$ in high-multiplicity collisions. To extend the study to even larger $\Delta\eta$, Fig.~\ref{fig:FTPCDphi} shows the TPC-FTPC $\Delta\phi$ correlations for associated particles in the Au and deuteron beam directions. Here the ZDC-Au energy is used for event selection to avoid self-correlations. There is a near-side long-range correlation in FTPC-Au at
$\Delta\eta\approx-3$ in high-multiplicity but not in low-multiplicity collisions. On the other hand, the TPC-FTPC correlations in the deuteron beam direction show no near-side peak. The physics origin of the observed near-side ridge at large $\Delta\eta$ in the Au beam direction is under investigation.

\section{Conclusions}
\label{sec:end}
Dihadron $\Delta\eta$-$\Delta\phi$ correlations in d+Au collisions at $\sqrt{s_{\text{NN}}} = 200$ GeV are presented. The multiplicity selection of events affects or biases the jet correlations. Simple subtraction of low-multiplicity data as a technique to remove jet contributions in high-multiplicity events at RHIC is problematic. A finite near-side correlated yield above a uniform background is observed at $|\Delta\eta|\approx1.5$  and $\Delta\eta\approx-3$ (in Au beam direction). 



%% References


%\bibliographystyle{elsarticle-num}

\begin{thebibliography}{00}

\bibitem{CMS1} V. Khachatryan {\it et al.} (CMS collaboration), J. High Energy Phys. {\bf 1009} (2010) 091.
\bibitem{CMS2} S. Chatrchyan {\it et al.} (CMS collaboration), Phys. Lett. B {\bf 718} (2013) 795814.
\bibitem{ALICE} B. Abelev {\it et al.} (ALICE collaboration), Phys. Lett. B {\bf 719} (2013) 29.
\bibitem{ATLAS} G. Ada {\it et al.} (ATLAS collaboration), Phys. Rev. Lett. {\bf 110} 182302 (2013).
\bibitem{PHENIX1} A. Adare {\it et al.} (PHENIX collabration), Phys. Rev. Lett. {\bf 111} 212301 (2013).
\bibitem{STARdAu} J. Adams {\it et al.} (STAR collaboration), Phys. Rev. C {\bf 72} 14904 (2005).
\bibitem{ALICEmassordering} B. Abelev {\it et al.} (ALICE collaboration), Phys. Lett. B {\bf 726} (2013) 164.
\bibitem{PHENIXmassordering} A. Adare {\it et al.}  (PHENIX collaboration), arXiv:1404.7461 [nucl-ex]
\bibitem{Bozek} P. Bozek and W. Broniowksi, Phys. Rev. C {\bf 88}, 014903 (2013).
\bibitem{Bzdak} A. Bzdak, B. Schenke, P. Tribedy, and R. Venugopalan, Phys. Rev. C {\bf 87}, 064906 (2013).
\bibitem{Qin} G.-Y. Qin and B. Mueller, Phys. Rev. C {\bf 89}, 044902 (2014).
\bibitem{Dusling} K. Dusling and R. Venugopalan, Phys. Rev. D {\bf 87}, 094034 (2013).
\bibitem{STAR:dAu4suppression} J. Adams,{\it et al.} (STAR collaboration), Phys. Rev. Lett. {\bf 91} 072304 (2003).
\bibitem{ZYAM} N. Ajitanand, J. Alxander, P. Chung, W. Holzmann, M. Issah, {\it et al.}, Phys. Rev. C {\bf 72} 011902 (2005).

\end{thebibliography}



\end{document}